\documentclass[journal=cmatex,manuscript=article]{achemso}

\usepackage[version=3]{mhchem} 
\usepackage{pslatex}
\usepackage{xr}
\usepackage{natbib}
\usepackage[labelfont=bf]{caption}
\usepackage[dvipsnames]{xcolor} 

\DeclareGraphicsExtensions{.pdf,.png,.jpg}
\graphicspath{{./Figures/}} 

\title{Mechanistically-guided materials chemistry: synthesis of new ternary nitrides, \ce{CaZrN2} and \ce{CaHfN2}}

\author{Christopher L. Rom}
\affiliation[Colorado State University]{Department of Chemistry, Colorado State University, Fort Collins, CO, USA}
\alsoaffiliation[National Renewable Energy Laboratory]{Materials Science Center, National Renewable Energy Laboratory, Golden, CO, USA}
\author{Andrew Novick}
\affiliation[Colorado School of Mines]{Department of Physics, Colorado School of Mines, Golden, CO, USA}
\author{Matthew J. McDermott}
\affiliation{Materials Sciences Division, Lawrence Berkeley National Laboratory, Berkeley, CA, USA}
\alsoaffiliation{Department of Materials Science and Engineering, University of California, Berkeley, CA, USA}
\author{Andrey A. Yakovenko}
\affiliation[Argonne National Laboratory]{X-ray Science Division, Advanced Photon Source, Argonne National Laboratory, Lemont, IL, USA}
\author{Jessica R. Gallawa}
\affiliation[Colorado State University]{Department of Chemistry, Colorado State University, Fort Collins, CO, USA}
\author{Gia Thinh Tran}
\affiliation[Colorado State University]{Department of Chemistry, Colorado State University, Fort Collins, CO, USA}
\author{Dominic C. Asebiah}
\affiliation[Colorado State University]{Department of Chemistry, Colorado State University, Fort Collins, CO, USA}
\author{Emily N. Storck}
\affiliation[Colorado State University]{Department of Chemistry, Colorado State University, Fort Collins, CO, USA}
\author{Brennan C. McBride}
\affiliation[Colorado State University]{Department of Chemistry, Colorado State University, Fort Collins, CO, USA}
\author{Rebecca C. Miller}
\affiliation[Colorado State University]{Analytical Resources Core, Colorado State University, Fort Collins, CO, USA}
\author{Amy L. Prieto}
\affiliation[Colorado State University]{Department of Chemistry, Colorado State University, Fort Collins, CO, USA}
\author{Kristin A. Persson}
\affiliation{Molecular Foundry, Lawrence Berkeley National Laboratory, Berkeley, CA, USA}
\alsoaffiliation{Department of Materials Science and Engineering, University of California, Berkeley, CA, USA}
\author{Eric Toberer}
\affiliation[Colorado School of Mines]{Department of Physics, Colorado School of Mines, Golden, CO, USA}
\author{Vladan Stevanović}
\affiliation[Colorado School of Mines]{Department of Physics, Colorado School of Mines, Golden, CO, USA}
\author{Andriy Zakutayev}
\affiliation[National Renewable Energy Laboratory]{Materials Science Center, National Renewable Energy Laboratory, Golden, CO, USA}
\author{James R. Neilson}
\altaffiliation{School of Advanced Materials Discovery, Colorado State University, Fort Collins, CO, USA}
\affiliation[Colorado State University]{Department of Chemistry, Colorado State University, Fort Collins, CO, USA}
\email{james.neilson@colostate.edu}

\begin{document}
\pagebreak

\begin{abstract}
Recent computational studies have predicted many new ternary nitrides, revealing synthetic opportunities in this underexplored phase space. 
However, synthesizing new ternary nitrides is difficult, in part because intermediate and product phases often have high cohesive energies that inhibit diffusion.
Here, we report the synthesis of two new phases, calcium zirconium nitride (\ce{CaZrN2}) and calcium hafnium nitride (\ce{CaHfN2}), by solid state metathesis reactions between \ce{Ca3N2} and \ce{$M$Cl4} ($M$ = Zr, Hf).  
Although the reaction nominally proceeds to the target phases in a 1:1 ratio of the precursors via \ce{Ca3N2 + $M$Cl4 -> Ca$M$N2 + 2 CaCl2}, reactions prepared this way result in Ca-poor materials (\ce{Ca_$x$ $M$_${2-x}$N2}, $x<1$). 
A small excess of \ce{Ca3N2} (ca. 20 mol\%) is needed to yield stoichiometric \ce{Ca$M$N2}, as confirmed by high-resolution synchrotron powder X-ray diffraction.
\textit{In situ} synchrotron X-ray diffraction studies reveal that nominally stoichiometric reactions produce \ce{Zr^{3+}} intermediates early in the reaction pathway, and the excess \ce{Ca3N2} is needed to reoxidize \ce{Zr^{3+}} intermediates back to the \ce{Zr^{4+}} oxidation state of \ce{CaZrN2}. 
Analysis of computationally-derived chemical potential diagrams rationalizes this synthetic approach and its contrast from the synthesis of \ce{MgZrN2}.
These findings additionally highlight the utility of \textit{in situ} diffraction studies and computational thermochemistry to provide mechanistic guidance for synthesis. 
\end{abstract}

\section{Introduction}

Predictive synthesis remains an important goal for materials chemists.\cite{kovnir2021predictive, neilson2023modernist, alberi2018_roadmap}
While many joint computational-experimental synthesis efforts have been reported in literature,\cite{sun2019map, walters2021first_predictive, gautier2015prediction, raccuglia2016machine, altman2020computationally, greenaway2022zinc, heinselman_thin_2019, hinuma2016discoveryNitridePredictions, gvozdetskyi2019computationally, zakutayev2013theoretical, woods2022roleZnZrN2, chen2022antimony, mcdermott2023assessing} trial-and-error approaches remain the dominant paradigm. 
Within this current paradigm, \textit{in situ} X-ray diffraction (XRD) is a powerful technique for materials discovery.\cite{shoemaker2014insitu, haynes2017panoramic, chen2022antimony, mcclain2021mechanistic, ito2021kinetically}
Sometimes called ``panoramic synthesis,'' \cite{shoemaker2014insitu, haynes2017panoramic, mcclain2021mechanistic} \textit{in situ} X-ray diffraction allows for observation of solid-state reactivity as it happens. 
Short-lived intermediate species and subtle crystallographic changes can be detected which would otherwise be missed by \textit{ex situ} experiments, accelerating materials discovery.
In a move towards predictive synthesis, \textit{in situ} XRD has been used in concert with computationally-generated chemical potential diagrams\cite{yokokawa1999generalizedChemicalPotentialDiagrams} to rationalize polymorph selectivity in the synthesis of ternary oxides\cite{todd2021selectivityPredominanceDiagrams}
However, this specific combined approach of \textit{in situ} XRD and chemical potential diagrams has yet to be used to discover new materials.

Nitrides are a compelling class of materials for melding computational guidance with \textit{in situ} X-ray diffraction (XRD) studies because they are difficult to synthesize.
High-temperatures are often needed to overcome the relative inertness of \ce{N2} (941 kJ/mol dissociation energy)\cite{CRC_2023} and drive solid state diffusion.\cite{greenaway2021ternaryReview, niewa1996groupReview}
However, elevated temperatures increase the entropic favorability of \ce{N2} gas, which leads many nitrides to decompose at relatively low temperatures.\cite{zakutayev2022experimentalSynthesisNitrides, elder1993thermodynamics}
Therefore, successful synthesises may be limited to narrow temperature windows, which \textit{in situ} XRD can help identify.\cite{rognerud2019kinetically} 
Recent computational studies have predicted numerous stable phases in previously uncharted spaces\cite{gharavi2018theoreticalPredictedMgMN2_Ti_Zr_Hf, hinuma2016discoveryNitridePredictions, sarmiento2015predictionOfPerovskiteNitrides, ching2001predictionOfSpinelNitrides, sun2019map}, accelerating the discovery of new phases (e.g., \ce{MgSnN2}\cite{greenaway2020combinatorialMgSnN2, kawamura2020synthesisMgSnN2_rocksalt}, \ce{CaSnN2}\cite{kawamura2021synthesisCaSnN2}, and \ce{MgZrN2}\cite{bauers2019ternaryrocksaltsemiconductors, rom2021bulk}).
Yet many predicted phases remain undiscovered, providing copious targets for synthesis.

\begin{figure}[ht!]
    \centering
    \includegraphics[width = 3.25in]{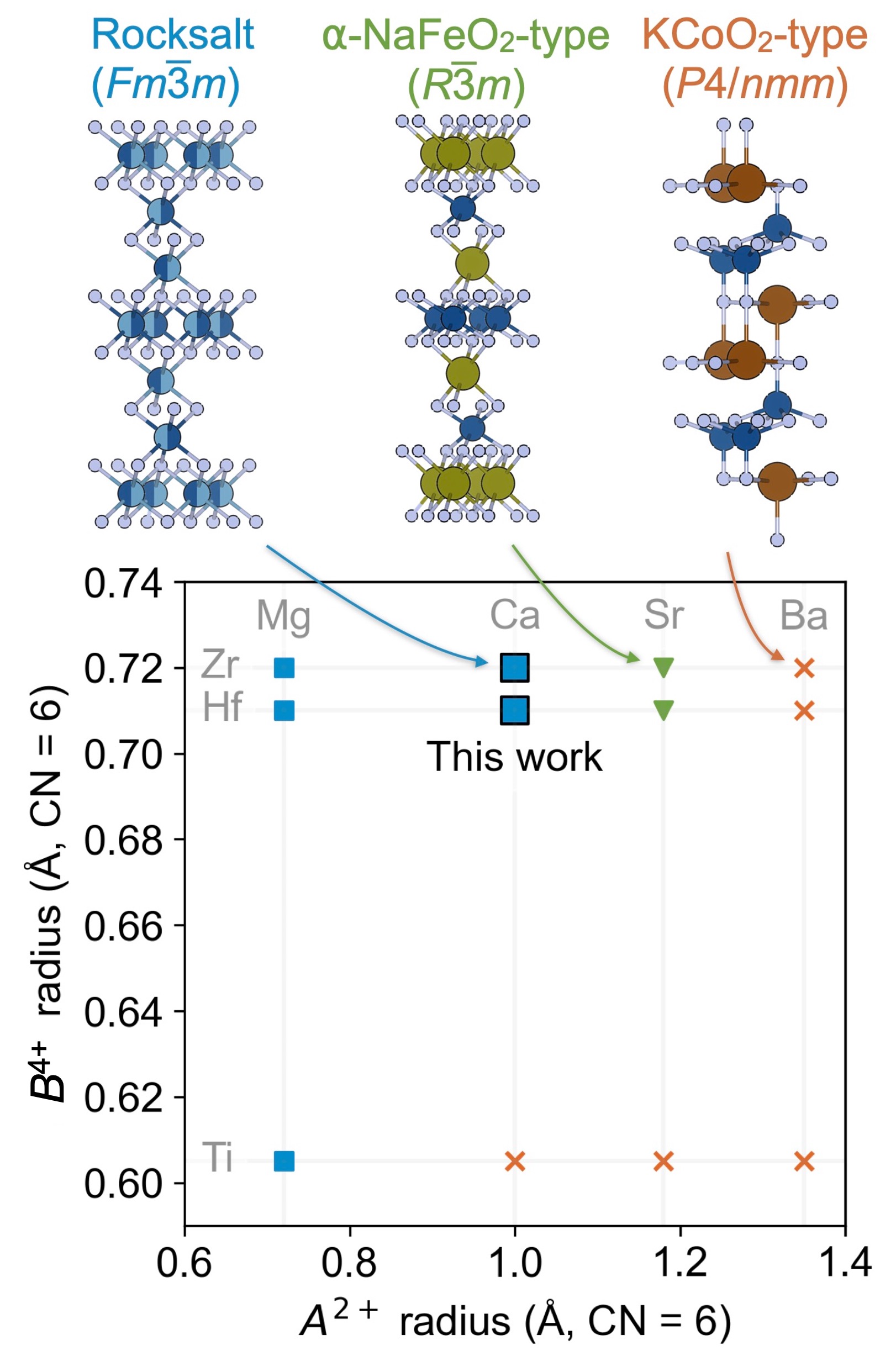}
    \caption{The \ce{$AB$N2} ternaries with larger alkali earth cations (Sr and Ba) crystallize in cation-ordered phases of the \ce{KCoO2}-type and \ce{$\alpha$-NaFeO2}-type structures, while \ce{Mg$B$N2} phases take the rocksalt (RS) structure. This work shows that \ce{CaZrN2} and \ce{CaHfN2} crystallize in the cation-disordered RS structure. Radii are sourced from literature\cite{shannon1969effective}.
    }
    \label{fig:context}
\end{figure}

All of the \ce{$AB$N2} phases ($A$ = Mg, Ca, Sr, Ba; $B$ = Ti, Zr, Hf) have been synthesized previously\cite{bauers2019ternaryrocksaltsemiconductors, li2017highCaTiN2, gregory1998synthesisSrTiN2, gregory1996synthesisSrZrN2_SrHfN2, shiraishi2022designBaTiN2}, with the exception of \ce{CaZrN2} and \ce{CaHfN2}\cite{zakutayev2022experimentalSynthesisNitrides}.
As summarized in Figure \ref{fig:context}, these phases crystallize with several structure types: rocksalt ($Fm\overline{3}m$), \ce{KCoO2}-type ($P4/nmm$), and \ce{$\alpha$-NaFeO2} ($R\overline{3}m$).
Given the intermediate size of \ce{Ca^{2+}} relative to the other alkali earths (Figure \ref{fig:context})\cite{shannon1969effective},  \ce{CaZrN2} and \ce{CaHfN2} were predicted to crystallize in the \ce{$\alpha$-NaFeO2} structure type\cite{orisakwe2014theoreticalAMN2}.
However, syntheses of \ce{CaZrN2} and \ce{CaHfN2} have not been previously reported, although one group described some synthetic attempts towards these ternaries that yielded only ZrN and HfN\cite{hunting2007synthesisCa4TiN4_Ca5NbN5}.

Here, we report the discovery of \ce{CaZrN2} and \ce{CaHfN2} in the disordered rocksalt (RS) structure via a metathesis approach. 
Heating \ce{Ca3N2} and \ce{ZrCl4} or \ce{HfCl4} up to a high-temperature dwell (ca. 1000~\textdegree{}C) for a brief period of time (10 min) yields \ce{CaZrN2} and \ce{CaHfN2} as fine powders (along with a byproduct \ce{CaCl2}). 
Importantly, a slight excess of \ce{Ca3N2} is needed, a fact we rationalize based on our \textit{in situ} synchrotron X-ray diffraction studies and computational thermodynamic analysis of the reaction pathway that show reduced \ce{Zr^{3+}} intermediates. 
Chemical potential diagrams rationalize the thermodynamics behind the observed reactions, accurately predicting differences in the reaction pathway between the reactions targeting \ce{CaZrN2} vs.\ \ce{MgZrN2} vs. \ce{ZnZrN2} (which has not yet been synthesized in bulk). 
Additional thermodynamic calculations show that these disordered RS ternaries are metastable (relative to the cation-ordered polymorph) even at the high synthesis temperatures. 
In sum, these findings show how chemical potential diagrams can augment \textit{in situ} XRD experiments and improve our ability to predict synthesis pathways.

\section{Results and Discussion}
\subsection{Structural and compositional analysis}
\begin{figure}[ht!]
\includegraphics[width=\textwidth]{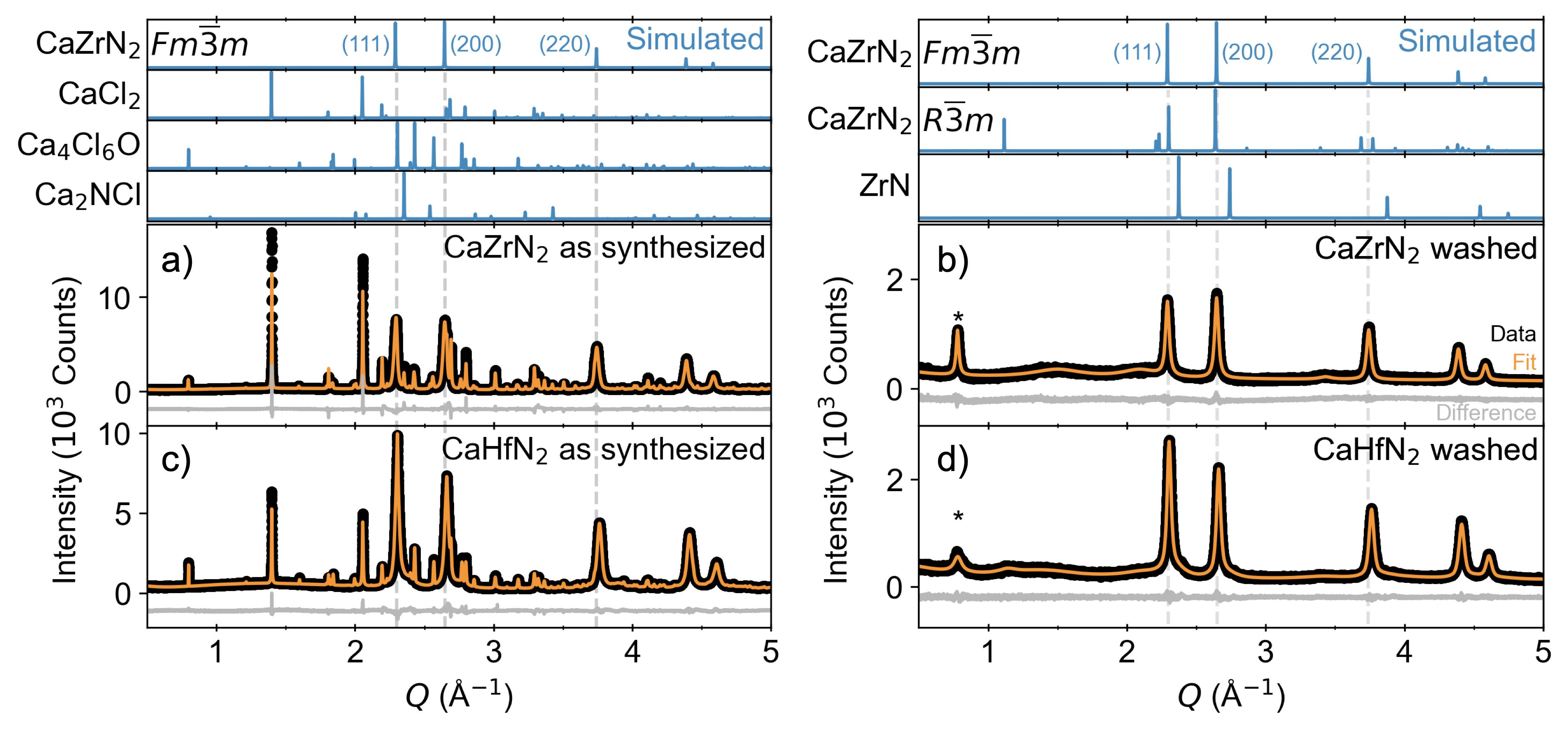}\\
\caption[SXRD patterns for washed samples of \ce{CaZrN2} and \ce{CaHfN2}]{
SXRD data of a) \ce{CaZrN2} as synthesized and b) washed along with c) \ce{CaHfN2} as synthesized and d) washed. 
Simulated patterns for relevant phases are shown for reference (top).
The predicted, ordered structure ($R\overline{3}m$) of \ce{Ca$M$N2} was not observed. 
A single impurity peak appears after washing (marked by $*$), possibly a disordered \ce{Ca4Cl6O} or calcium methoxide.
Dashed gray vertical lines guide the eye to the (111), (200), and (220) reflections of \ce{Ca$M$N2} $Fm\overline{3}m$.
Full diffraction patterns are shown in Figure S2,  
and refinement results are in Tables S1-S2 and Figure S3.  
}
  \label{fig:main_CaZrN2_XRD} 
\end{figure}

High-resolution synchrotron powder X-ray diffraction (SXRD) measurements show that metathesis reactions between \ce{Ca3N2} and \ce{$M$Cl4} ($M$ = Zr, Hf) yielded \ce{CaZrN2} and \ce{CaHfN2} in the RS structure ($Fm\bar{3}m$, Figure \ref{fig:main_CaZrN2_XRD}). 
These samples (\ce{1.21\ Ca3N2 + ZrCl4} and \ce{1.16\ Ca3N2 + HfCl4}) were heated at +5~\textdegree{}C/min to 1100~\textdegree{}C, dwelled for 10 min, and then air-quenched to yield \ce{Ca$M$N2} along with an array of side products (\ce{CaCl2}, \ce{Ca2NCl}, \ce{Ca4Cl6O}, and in the case of $M$ = Hf, trace HfN). 
Recovered samples were black, red, or green depending on precise synthesis conditions (Figure S1).  
Washing the samples with anhydrous methanol removes the chloride-containing byproducts (\ce{CaCl2}, \ce{Ca2NCl}, \ce{Ca4Cl6O}).  
However, one unindexed diffraction peak is present at $Q=0.8$~\AA{}$^{-1}$.
This peak matches both the (100) reflection from \ce{Ca4Cl6O} and the most intense reflection in a prior report of calcium methoxide\cite{masood2012synthesis_calcium_methoxide}.
Unfortunately, the presence of the impurity phase inhibits accurate compositional analysis by Scanning Electron Microscopy (SEM) with Energy Dispersive X-ray Spectroscopy (EDS), rigorously air-free X-ray Photoelectron Spectroscopy (XPS), and combustion analysis (further described in the Supporting Information, Figures S5-S8  
and Tables S3-S4).\cite{schneider2020designAirFreeXPS, fairley2021systematicCASAXPS, coelho2018topas, azdad2018valenceXPS} 
Consequently, our characterization efforts centered on diffraction. 

Rietveld analysis of the RS products show these phases are consistent with cation-disordered \ce{CaZrN2} and \ce{CaHfN2}.
The metal site occupancy ($x$ in \ce{Ca_$x$ $M$_{$2-x$}N2}) refined to 1.00(2) for the washed \ce{CaZrN2} and 1.00(1) for the washed \ce{CaHfN2} sample.
Washing does not substantially change the refined metal site occupancy or lattice parameter compared to the samples as synthesized (Figure S3).  
Lattice parameters of the washed \ce{Ca$M$N2} phases refined to substantially larger unit values ($a=4.7486(3)$~\AA{} and $4.7224(4)$~\AA{} for \ce{CaZrN2} and \ce{CaHfN2}, respectively) than their isostructural binaries ($a = 4.58$~\AA{} and $4.52$~\AA{} for ZrN and HfN, respectively)\cite{christensen_neutron_1975, aigner1994latticeThermalExpansionZrN_HfN}.
These unit cell values are close to the Ca-N-Zr distance in the computationally-predicted, cation-ordered \ce{CaZrN2} structure (4.744~\AA{}), which serves as an estimate for the theoretical disordered RS lattice parameter (Figure S4).  
We focused most of our work on the Zr analog, since Zr and Hf exhibit analogous chemistry in this case. 

\begin{figure}[ht!]
    \centering
    \includegraphics[width=0.5\textwidth]{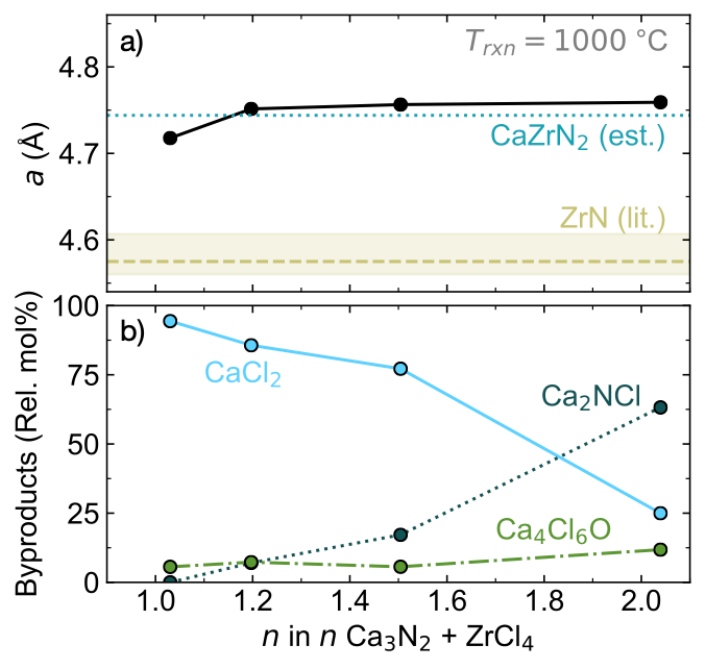}
    \caption[Effect of precursor ratio $n$ in \ce{$n$Ca3N2 + ZrCl4} on a) the lattice parameter of \ce{CaZrN2} and byproducts of synthesis]{a) The precursor ratio (expressed as $n$ in \ce{$n$Ca3N2 + ZrCl4}) impacted the lattice parameter of the RS phase and b) the distribution of byproduct phases, as determined by Rietveld analysis. 
    Approximately 20\% mole excess ($n=1.2$) \ce{Ca3N2} was necessary to yield stoichiometric \ce{CaZrN2}, as indicated by the estimated value of $a = 4.744$~\AA{} from the computationally-predicted structure (Figure S4).
    However, some of the excess \ce{Ca3N2} also yielded a \ce{Ca2NCl} byproduct by reacting with \ce{CaCl2} (bottom). 
    The byproduct relative mol\% of \ce{CaCl2}, \ce{Ca2NCl}, and \ce{Ca4Cl6O} sum to 100\% (i.e., \ce{Ca_${x}$Zr_${2-x}$N2} is not included owing to uncertainty in $x$).
    These data were from samples held in alumina crucibles sealed in ampules under vacuum and heated at 5~\textdegree{}C/min to 1000~°C, dwelled for 10~min, then quenched by removing the ampule from the furnace and placing on the benchtop.}
    \label{fig:ratio}
\end{figure}

A slight excess of \ce{Ca3N2} (ca. 20 mol\%) is required to yield stoichiometric \ce{CaZrN2} (Figure \ref{fig:ratio}), as measured by the product lattice parameter. 
Although the nominal balanced equation for the reaction is \ce{Ca3N2 + ZrCl4 -> CaZrN2 + 2\ CaCl2}, a 1:1 precursor ratio yields a lattice parameter of 4.72~\AA{}, trending towards ZrN and suggesting a Ca-poor phase. 
With an additional 20 mol\% \ce{Ca3N2} (i.e., $n=1.20$), the RS lattice parameter is closer to the 4.744~\AA{} value estimated from the computationally-predicted structure (Figure S4).  
The continued growth in the RS lattice parameter above $n=1.20$ could suggest the formation of Ca-interstitials, but characterizing such defects is beyond the scope of this study.
This trend contrasts with our prior synthesis of \ce{MgZrN2}, where the stoichiometric reaction \ce{2.0\ Mg2NCl + ZrCl4 -> MgZrN2 + 3\ MgCl2} produced the targeted ternary (RS \ce{MgZrN2}), and additional \ce{Mg2NCl} did not change the RS lattice parameter or increase Mg content (as measured by SXRD, Inductively Coupled Plasma  Atomic Emission Spectroscopy, and Energy Dispersive X-ray Spectroscopy).\cite{rom2021bulk}
Some of the excess \ce{Ca3N2} reacts with the \ce{CaCl2} byproduct to make \ce{Ca2NCl}, which is observed in an increasing mol\% with increasing \ce{Ca3N2} content.
The need of excess \ce{Ca3N2} is not attributable to the oxide impurity in the \ce{Ca3N2} precursor, which is estimated at $<5$ wt\% of our precursor by quantitative Rietveld analysis (below the detection limit of laboratory PXRD). 
We also note that increasing dwell time from 10~min to 4~h did not substantially change the RS lattice parameter, size, or strain for Ca-rich reactions (\ce{2.0\ Ca3N2 + ZrCl4}, $T_\mathrm{rxn} = 1000$~\textdegree{}C, Figure S10).  
Increasing temperature decreases the lattice parameter but increases the crystallite size (\ce{1.2\ Ca3N2 + ZrCl4}, dwell time = 10~min, Figure S11). 
Furthermore, increasing dwell times for Ca-poor reactions decreases the rocksalt lattice parameter (dwell temperature = 900~\textdegree{}C, Figure S12).
The key observation is that some excess \ce{Ca3N2} is necessary. 
This result is explained by observations from our \textit{in situ} SXRD measurements, as discussed in the next section.

Using metathesis (ion exchange) reactions is also a key synthetic choice for the synthesis of \ce{CaZrN2}, as this process circumvents the solid-state diffusion challenges common to traditional ceramic syntheses\cite{martinolich2016circumventingDiffusion, gillan1996synthesisRefractory}.
A control reaction between \ce{Ca3N2} and \ce{Zr} under flowing \ce{N2} at 1000~\textdegree{}C for 10~h produces only \ce{ZrN} alongside unreacted \ce{Ca3N2}, indicating that Ca diffusion into the ZrN is slow (Figure S13).  
Literature also shows the difficulty of making \ce{CaZrN2} and \ce{CaHfN2}. 
In their report on the synthesis of \ce{Ca4TiN4} and \ce{Ca5NbN5} with the assistance of a \ce{Li3N} flux, Hunting,~et al., noted that similar reactions between \ce{Ca3N2} and \ce{Zr} or \ce{Hf} powders in \ce{Li3N} flux only yielded the ZrN and HfN\cite{hunting2007synthesisCa4TiN4_Ca5NbN5}. 
And although many closely related compounds have been synthesized by ceramic techniques (i.e., \ce{CaTiN2}, \ce{SrZrN2}, \ce{BaZrN2})\cite{li2017highCaTiN2, gregory1996synthesisSrZrN2_SrHfN2, gregory1998synthesis_AMN2_incl_BaZrN2}, the absence of \ce{CaZrN2} and \ce{CaHfN2} from the literature implies that a synthetic strategy had previously been elusive.

We acknowledge that the \textit{perfectly stoichiometric} compounds \ce{CaZrN2} and \ce{CaHfN2} may not be synthesized from the reactions, \ce{1.21\ Ca3N2 + ZrCl4} and \ce{1.16\ Ca3N2 + HfCl4}; however, these were the optimal reaction conditions identified in this study. These synthesis conditions produced RS structures most consistent with the title compounds, given our structural analysis of lattice parameter and cation site occupancy. Detailed compositional analysis was not possible with our current samples given an unidentified impurity phase (See the Supporting Information, Figures S5-S8 and Tables S3-S4). Precise analysis of composition (including defects) will be essential for better understanding the potential applications of these new semiconductors, but such analysis is beyond the scope of this study. Instead, we focus on the mechanisms of this synthesis, with the goal of showing how \textit{in situ} SXRD studies and thermodynamic calculations can guide metathesis reactions for the synthesis of elusive new materials. 

\subsection{\textit{In situ} synchrotron powder X-ray diffraction}

\begin{figure}[ht!]
\includegraphics[width=\textwidth]{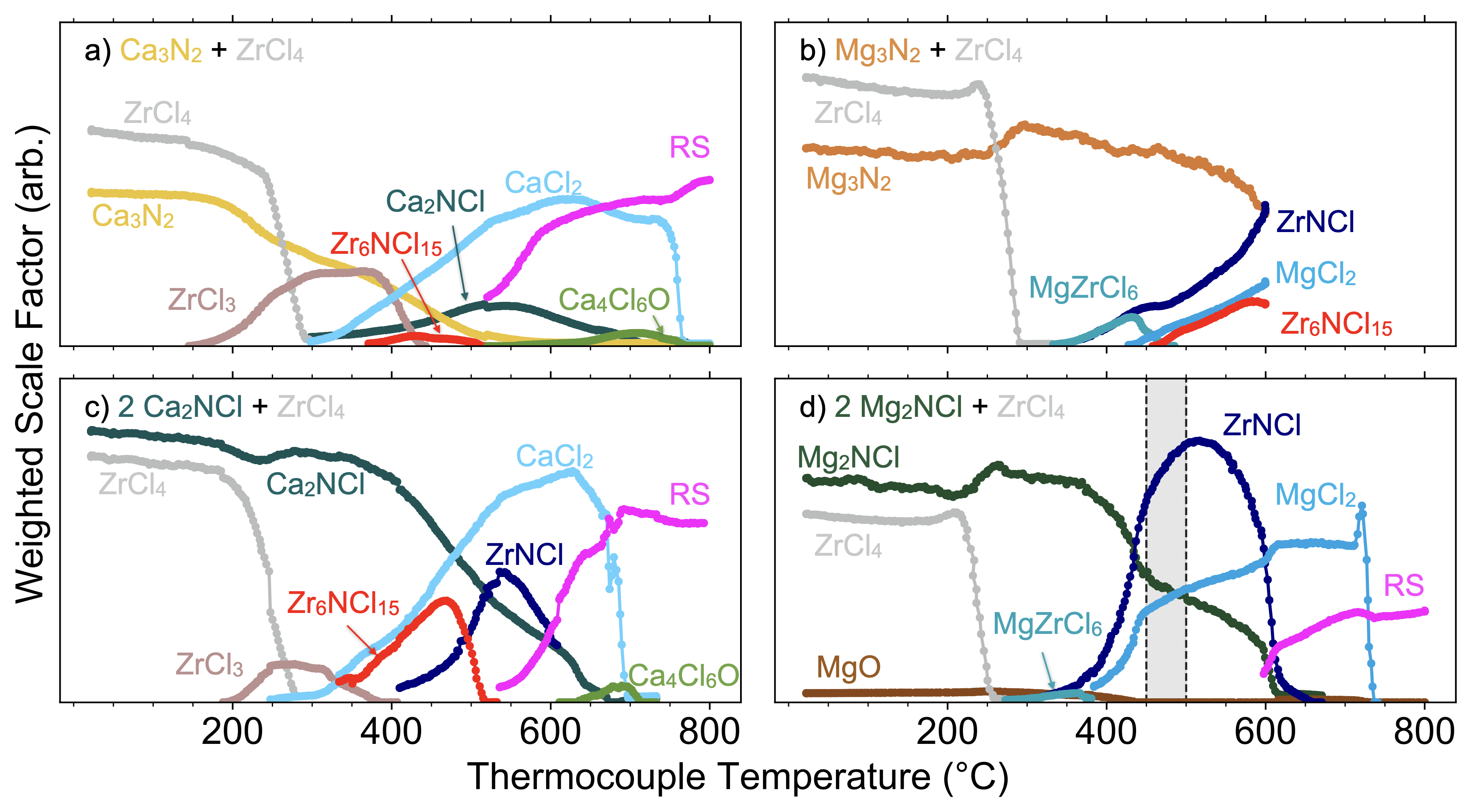}\\
  \caption[Rietveld analysis of \textit{in situ} SXRD data on metathesis reactions targeting \ce{MgZrN2} and \ce{CaZrN2}]{Phase contributions to the \textit{in situ} SXRD data as a function of temperature as indicated by the weighted scale factors (W.S.F.) for reactants, intermediates, and products upon heating precursor mixtures of a) \ce{Ca3N2 + ZrCl4}, b) \ce{Mg3N2 + ZrCl4}, c) \ce{2\ Ca2NCl + ZrCl4}, and d) \ce{2\ Mg2NCl + ZrCl4}. 
  In reactions with \ce{Mg3N2}, \ce{Ca3N2}, and \ce{Ca2NCl}, trivalent Zr species were observed (\ce{ZrCl3} and \ce{Zr6NCl15}), indicating reduction of \ce{Zr^{4+}} to \ce{Zr^{3+}} and loss of \ce{N2}. Only the reaction using \ce{Mg2NCl} maintained the tetravalent oxidation state of \ce{Zr}. The ternary phases \ce{Ca_$x$Zr_${2-x}$N2} and \ce{Mg_$x$Zr_${2-x}$N2} are denoted as ``RS''. The grey highlighted region of (d) shows the temperature range used for step 1 of the two-step synthesis of \ce{MgZrN2} detailed in prior reports\cite{rom2021bulk, todd2021twostep}.
  }
  \label{fig:insituWSF}
\end{figure}
\textit{In situ} synchrotron powder X-ray diffraction studies (SXRD) at the Advanced Photon Source (17-BM) reveals the pathway through which the metathesis reactions proceed. We examined reactions in sealed quartz capillaries targeting \ce{CaZrN2} (and \ce{MgZrN2}, for comparison) starting from reagent mixtures for the ideal reactions: \ce{$A$3N2 + ZrCl4 -> $A$ZrN2 + 2\ $A$Cl2} and \ce{2\ $A$2NCl + ZrCl4 -> $A$ZrN2 + 3\ $A$Cl2} ($A$ = Mg, Ca). These ratios are calcium-deficient compared to the optimized synthesis of \ce{CaZrN2} discussed previously, yet are nonetheless illuminating. They show how ostensibly similar precursors (\ce{$A$3N2} and \ce{$A$2NCl}) proceed via different reaction pathways. These differences in reaction pathway ultimately required different synthetic conditions for the optimized syntheses of \ce{MgZrN2} and \ce{CaZrN2}.  The relative amounts of crystalline phases present during the reaction (determined by quantitative phase analysis using the Rietveld method of \textit{in situ} diffraction data) are shown in Figure \ref{fig:insituWSF}, with the raw diffraction patterns shown in Figures S17 and S18. The observed Zr-containing intermediates are summarized in Table \ref{tab:insitu_Zr_summary}.

Quantitative phase analysis of \textit{in situ} SXRD data from the reaction between \ce{Ca3N2 + ZrCl4} show that reduced \ce{Zr^{3+}} species form at low reaction temperatures (Figure \ref{fig:insituWSF}a). \ce{ZrCl3} is the first intermediate to form, appearing by 200~\textdegree{}C. 
Subsequently, \ce{ZrCl4} disappears from the patterns as it sublimes ($T_\mathrm{sub}$ = 331~\textdegree{}C), at which point \ce{CaCl2} begins to grow in. 
Near 400~\textdegree{}C, the \ce{ZrCl3} converts to a nitride chloride, \ce{Zr6NCl15}. 
Throughout this process, the \ce{Ca3N2} precursor steadily declines. 
\ce{Ca3N2} is fully consumed by 600~\textdegree{}C, with some of it reacting with the \ce{CaCl2} byproduct to make \ce{Ca2NCl} (\ce{Ca3N2 + CaCl2 -> 2\ Ca2NCl}).
\ce{Ca2NCl} persists up to 700~\textdegree{}C. 
By ca. 550~\textdegree{}C, the RS phase \ce{Ca_$x$Zr_${2-x}$N2} begins to grow in (initially as ZrN, as will be detailed subsequently). 
A minor phase of \ce{Ca4Cl6O} also grows in, indicating a small degree of oxygen impurity. 
The reaction of \ce{2\ Ca2NCl + ZrCl4} proceeds in a similar fashion as \ce{Ca3N2 + ZrCl4}. 
However, using \ce{Ca2NCl} also includes a ZrNCl-like intermediate, which is not detected in the \ce{Ca3N2} reaction (Figure S17).  
In sum, the initial reaction between \ce{Ca3N2 + ZrCl4} proceeds through \ce{Zr^{3+}} intermediates via the following balanced equations:
\begin{align}
 \ce{Ca3N2 + 6\ ZrCl4 &-> 6\ ZrCl3 + N2 + 3\ CaCl2}, \label{eq:ZrCl4_ZrCl3}\\ 
 \ce{Ca3N2 + 12\ ZrCl3 &-> 2\ Zr6NCl15 + 3\ CaCl2}, \label{eq:ZrCl3_Zr6NCl15}\\ 
 \ce{5\ Ca3N2 + 2\ Zr6NCl15 &-> 12\ ZrN + 15\ CaCl2}. \label{eq:Zr6NCl15_ZrN} 
\end{align}
The final step of this process involves the reaction of the \ce{Ca2NCl} intermediate with the \ce{ZrN} rocksalt phase:
\begin{equation}
     \ce{4\ Ca2NCl + 6\ ZrN + N2 -> 6\ CaZrN2 + 2\ CaCl2}. \label{eq:ZrN_CaZrN2} 
\end{equation}

In contrast, the analysis of \textit{in situ} SXRD data in the analogous Mg system shows reactivity differences between \ce{Mg3N2} and \ce{Mg2NCl}, with \ce{Mg2NCl} conserving the tetravalent state of \ce{Zr^{4+}} throughout the reaction pathway (Figure \ref{fig:insituWSF}b and d, Table \ref{tab:insitu_Zr_summary}). 
The observed reaction pathways are consistent with our prior (lower resolution) \textit{in situ} study on the synthesis of \ce{MgZrN2}.\cite{rom2021bulk}
Both the reaction of \ce{Mg3N2 + ZrCl4} and the reaction of \ce{2 Mg2NCl + ZrCl4} begin with \ce{ZrCl4} sublimation, followed by the formation of a small amount of \ce{MgZrCl6} near 300~\textdegree{}C. 
However, the reaction with \ce{Mg3N2} goes on to produce \ce{Zr6NCl15} near 430~\textdegree{}C, which indicates the reduction of \ce{Zr^{4+}} to \ce{Zr^{3+}}, likely paired with the production of \ce{N2}. 
Furthermore, the amount of \ce{Mg3N2} only decreases slightly by 600~\textdegree{}C (at which point, the measurement was stopped). 
This result suggests that \ce{Mg3N2} is a kinetically slow precursor. 
On the other hand, the \ce{Mg2NCl} reaction does not produce any distinct \ce{Zr^{3+}} intermediates. 
Instead, the amount of \ce{Mg2NCl} decreases in a stepwise fashion near 440~\textdegree{}C coincident with the formation of \ce{ZrNCl} and the rapid increase of \ce{MgCl2}. 
A second stepwise decrease in \ce{Mg2NCl} occurs near 600~\textdegree{}C, coincident with the consumption of the \ce{ZrNCl} and the formation of the RS phase (\ce{Mg_$x$Zr_${2-x}$N2}). 
Therefore, the synthesis of \ce{MgZrN2} appears to proceed through two reaction steps:
\begin{align}
\ce{Mg2NCl + ZrCl4 &-> ZrNCl + 2\ MgCl2}, \label{eq:Mg2NCl_ZrNCl} \\  
\ce{Mg2NCl + ZrNCl &-> MgZrN2 + MgCl2}. \label{eq:ZrNCl_MgZrN2}  
\end{align}
As noted in prior reports, heating directly to a high temperature (ca. 800~\textdegree{}C) results in an Mg-poor RS phase, but a two-step process yields stoichiometric \ce{MgZrN2}\cite{rom2021bulk,todd2021twostep}. 
The first step of that two-step process ($T_{react}$) is a 12 to 24 h dwell at 450--500~\textdegree{}C (the highlighted grey temperature region in Figure \ref{fig:insituWSF}), likely where Equations \ref{eq:Mg2NCl_ZrNCl} and \ref{eq:ZrNCl_MgZrN2} proceed to completion.
The second heating step (to $T_{cryst}=800$~\textdegree{}C) aids in the crystallization of the amorphous/nanocrystalline \ce{MgZrN2}.
This \textit{in situ} study reveals how the two-step process allows the synthesis of \ce{MgZrN2} to proceed through a single \ce{Zr^{4+}} intermediate, whereas the synthesis of \ce{CaZrN2} must proceed through \ce{Zr^{3+}} intermediates that require subsequent re-oxidation (Table \ref{tab:insitu_Zr_summary}).

\begin{table}[ht!]
    \centering
\caption{Summary of Zr-containing intermediates via the \textit{in situ} SXRD studies shown in Figure \ref{fig:insituWSF}}
\label{tab:insitu_Zr_summary}
    \begin{tabular}{c|cccc}
         Reaction&   \multicolumn{4}{c}{Zr-containing intermediates}  \\ \hline
         \ce{Ca3N2 + ZrCl4}&  \ce{ZrCl3}&  \ce{Zr6NCl15}&  &  \\
         \ce{2\ Ca2NCl + ZrCl4}&  \ce{ZrCl3}&  \ce{Zr6NCl15}&  ZrNCl&  \\
         \ce{Mg3N2 + ZrCl4}&  &  \ce{Zr6NCl15}&  ZrNCl&  \ce{MgZrCl6}\\
         \ce{2\ Mg2NCl + ZrCl4} &  &  &  ZrNCl&  \ce{MgZrCl6}\\ \hline
    \end{tabular}
\end{table}

\begin{figure}[ht!]
    \centering
    \includegraphics[width = 3.25in]{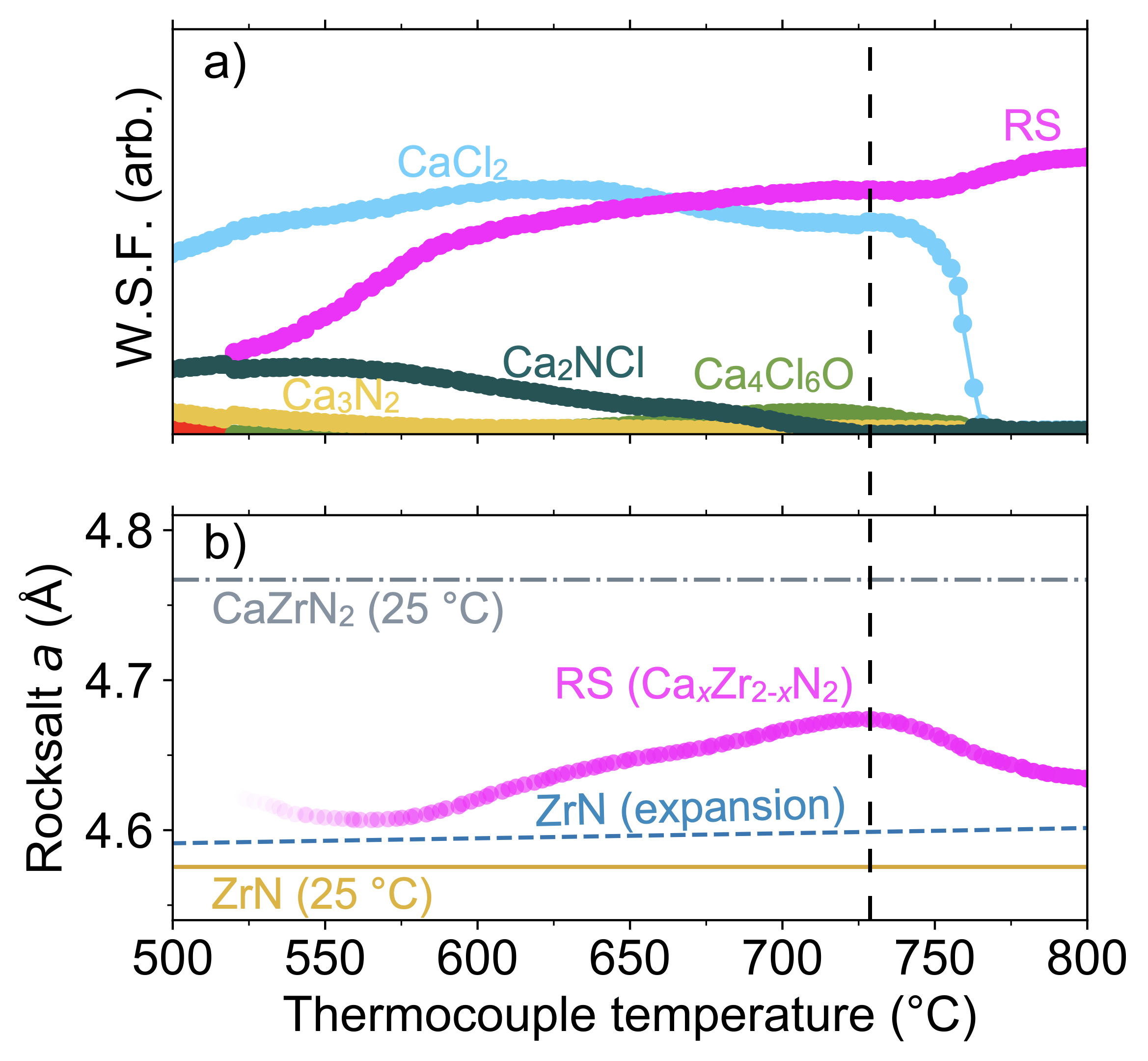}
    \caption{a) Weighted scale factor (W.S.F.)  as a function of temperature for the \ce{Ca3N2 + ZrCl4} reaction and b) refined lattice parameter $a$ for the RS \ce{Ca_$x$Zr_$2-x$N2} obtained from Rietveld analysis of the \textit{in situ} SXRD data. The opacity of the RS markers in (b) is normalized to the W.S.F. in (a). The lattice parameters for ZrN and \ce{CaZrN2} at room temperature are shown for reference, along with the value of ZrN accounting for thermal expansion\cite{houska1964thermalExpansion}. The increase in $a$ for \ce{Ca_$x$Zr_$2-x$N2} suggests Ca uptake from the \ce{Ca2NCl} intermediate. When the \ce{Ca2NCl} intermediate is fully consumed (vertical dashed line), the RS lattice parameter begins contracting, suggesting Ca loss.}
    \label{fig:wsf_v_lattice}
\end{figure}

The lattice parameter, $a$, of the disordered rocksalt \ce{Ca_$x$Zr_$2-x$N2} formed during  the \ce{Ca3N2 + ZrCl4} reaction shows that the ZrN intermediate from Equation \ref{eq:Zr6NCl15_ZrN} reacts with \ce{Ca2NCl} to form \ce{CaZrN2} (Figure \ref{fig:wsf_v_lattice}). 
As the RS phase grows in between 500 and 600~\textdegree{}C, the lattice parameter closely matches the value expected for ZrN by thermal expansion\cite{houska1964thermalExpansion}. 
Between 520~\textdegree{}C and 570~\textdegree{}C, the apparent decrease in the parameter is an artifact of the broad, low-intensity peaks that are difficult to accurately fit. 
Above 570\textdegree{}C, the parameter expands up to 4.674~\AA{} by 725~\textdegree{}C.
The expansion concomitant with the decreasing W.S.F. of \ce{Ca2NCl} suggests calcium uptake by ZrN from the \ce{Ca2NCl} intermediate across a solid solution \ce{Ca_$x$ Zr_${2-x}$ N2} ($0\leq x \leq 1$), with \ce{Zr^{3+}} to \ce{Zr^{4+}} oxidation driven by  \ce{N2} reincorporation:
\begin{equation}
\ce{$4x$\ Ca2NCl + (12-6$x$)\ ZrN + $x$\ N2 -> 6\ Ca_$x$Zr_{2-$x$}N2 + $2x$\  CaCl2}\label{eq:N2_hypothesis_ss}  
\end{equation}
Above 725~\textdegree{}C, the RS lattice parameter decreases.
This inflection point occurs at the same temperature at which \ce{Ca2NCl} is fully consumed. 
The contraction of the lattice parameter is consistent with a loss of calcium from the structure, possibly by reaction with oxygen from the quartz container (\ce{2\ Ca_$x$Zr_${2-x}$N2 + $x$\  SiO2 -> $2x$\ CaO + $(4-2x)$\ ZrN + $x$\ N2 + $x$\ Si}).

This pathway shows why ``excess'' \ce{Ca3N2} is required for \ce{CaZrN2} formation: it provides the necessary oxidant, \ce{N2}. 
The \ce{N2} likely comes from the reincorporation of the released nitrogen yielded during \ce{Zr^{3+}} formation (Equation \ref{eq:ZrCl4_ZrCl3}), although alternative hypotheses are considered in the Supporting Information (Equations S1 and S2).  
While this process may seem similar to the ceramic reaction discussed previously (Figure S13),  
the ZrN crystallites formed \textit{in situ} via metathesis are very small (ca. 10~nm) compared to the ZrN formed via the ceramic route (ca. 1~$\mu$m) and may contain a significant fraction of point defects, facilitating diffusion. 
These differences allow the metathesis-produced ZrN to react to form \ce{CaZrN2} within a 10~min dwell while the ceramic route is impractically slow.
To test the Equation \ref{eq:N2_hypothesis_ss} hypothesis, we reacted \ce{2\ Ca3N2 + ZrCl4} at 1000~\textdegree{}C under flowing \ce{N2} for 4~h (Figure S13).  
The resulting PXRD data of the red powder was well described by rocksalt \ce{CaZrN2} ($a = 4.764$~\AA{}) with \ce{Ca2NCl} as the only byproduct.
Furthermore, thermodynamic calculations (see next section) suggest Equation \ref{eq:N2_hypothesis_ss} is a thermodynamically favorable step.

\subsection{Thermodynamic analysis of the reaction pathway}

\begin{figure}[ht!]
\includegraphics[width= 3.25 in]{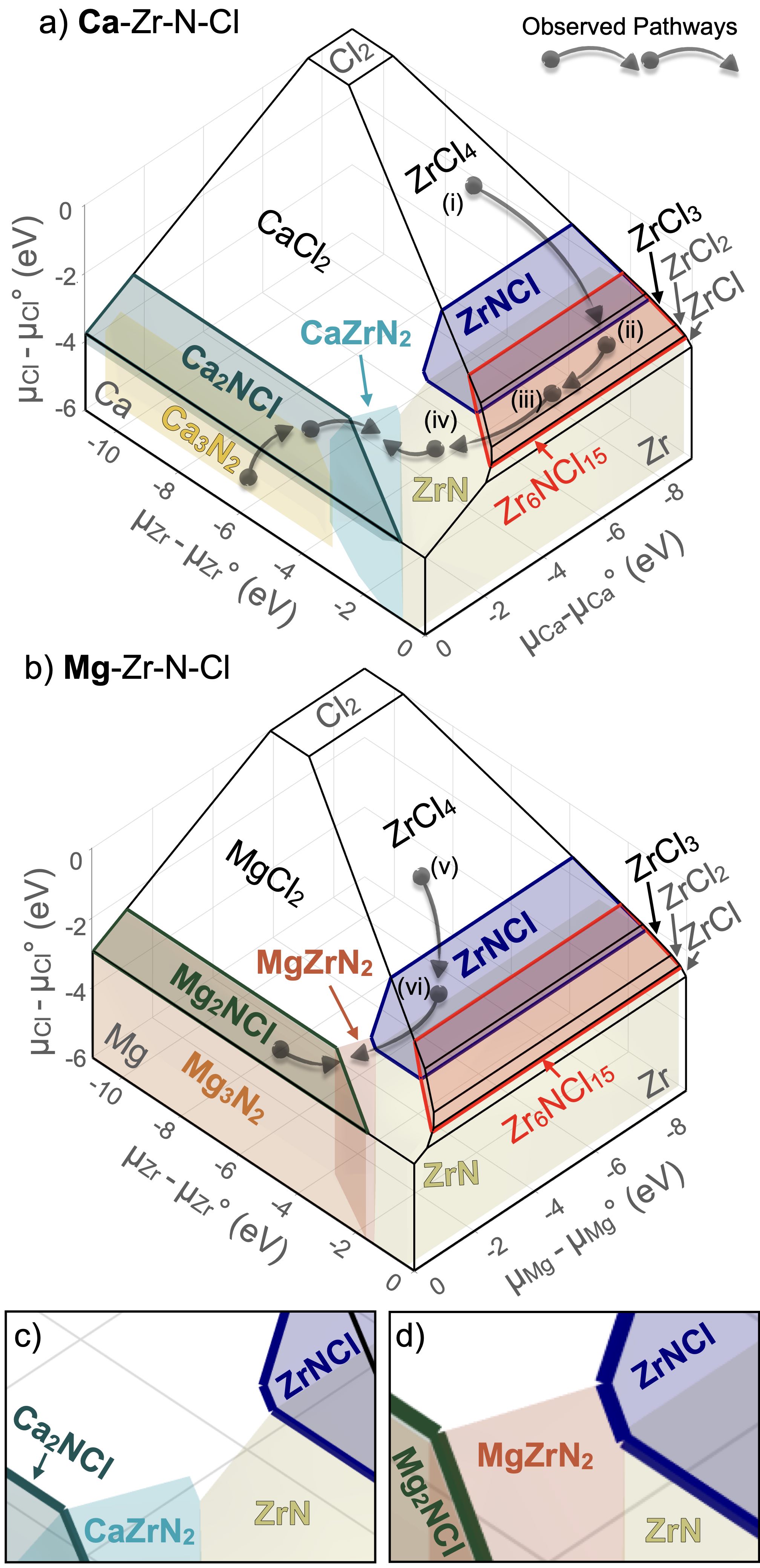}\\
\caption{Chemical potential diagrams for a) the Ca-Zr-N-Cl chemical system and b) the Mg-Zr-N-Cl chemical system calculated for 300~\textdegree{}C. Focused view highlighting the interfaces between \ce{ZrNCl} and c) \ce{Ca2NCl} or d) \ce{Mg2NCl}.
Select nitrogen-containing phases in the 4-component system are illustrated via their intersection with the 3-dimensional $\mu_\mathrm{Ca}-\mu_\mathrm{Zr}-\mu_\mathrm{Cl}$ or $\mu_\mathrm{Mg}-\mu_\mathrm{Zr}-\mu_\mathrm{Cl}$ spaces. 
These intersections may appear with complicated volumes due to the lower dimensional intersection (e.g., \ce{CaZrN2}, \ce{ZrN}). Full quaternary diagrams are shown in Figure S22.
Curved arrows qualitatively illustrate the experimentally observed reaction pathways.
}
  \label{fig:predominance}
\end{figure}

The chemical potential diagrams shown in Figure \ref{fig:predominance} describe the energetic landscape through which these metathesis reactions proceed. 
These models are computed using calculated phase energies from the Materials Project database\cite{jain2013commentaryMaterialsProject} (see Methods).
Each visible facet is a stable phase in the $A$-Zr-Cl system; larger facets indicate phases with deeper stability on the convex hull of a compositional phase diagram relative to competing phases. Nitrogen-containing phases are illustrated by 3-dimensional ($A$-Zr-Cl) polyhedral slices of the full 4-dimensional polytopes.  
Intersection points, edges, or faces between these regions of chemical potential space indicate which phases share stable interfaces.
It has previously been observed that a chemical reaction proceeding under local equilibrium (i.e., diffusion-controlled) conditions will follow a pathway between adjacent phases\cite{todd2021selectivityPredominanceDiagrams, neilson2023modernist}.

Figure \ref{fig:predominance}a and c show that no \ce{Zr^{4+}}-containing species in the Ca-Zr-N-Cl phase space shares a stable interface with the \ce{CaZrN2} product, indicating that \ce{Zr^{3+}}-containing intermediates must form if the system proceeds through local equilibrium at interfaces. Specifically, the region of stability for ZrN connects both the \ce{Zr6NCl15} and \ce{ZrNCl} spaces to the \ce{CaZrN2} space, indicating that ZrN should form along the pathway to \ce{CaZrN2}. In reality, due to the accommodation of off-stoichiometry, there is likely no sharp boundary (discontinuity) between ZrN and \ce{CaZrN2}.\cite{yokokawa1999generalizedChemicalPotentialDiagrams} Rather, this sharp transition is due to the limitations of our thermodynamic data, as only ordered, stoichiometric compounds are permitted in the DFT calculations of the Materials Project database. We discuss the stability of cation-disordered \ce{CaZrN2}  further in the next section and in the Supporting Information (Figure S21).

The chemical potential diagram for the Mg-Zr-N-Cl system (Figure \ref{fig:predominance}b) appears similar to the Ca system but with one key difference: ZrNCl shares a stable interface with both the precursor \ce{ZrCl4} and the product \ce{MgZrN2} (Figure \ref{fig:predominance}d). This shared boundary suggests that the intermediate ZrNCl (a \ce{Zr^{4+}}-containing phase) facilitates the formation of \ce{MgZrN2} at the \ce{ZrCl4}$\vert$\ce{Mg2NCl} interface while preventing the system from undergoing Zr and N-based reduction and oxidation. 

Reaction network analysis\cite{mcdermott2021graph} can quantify these differences between the Ca- and Mg-Zr-N-Cl systems. The method calculates a distance in chemical potential space, simplifying these geometric representations into numeric values. The chemical potential distance for the reaction of \ce{Ca2NCl + ZrNCl -> CaZrN2 + CaCl2} is 0.733 eV/atom, where the non-zero number means that the reactants and products do not share stable interfaces (as visualized in Figure \ref{fig:predominance}c). In contrast, \ce{Mg2NCl + ZrNCl -> MgZrN2 + MgCl2} (Figure \ref{fig:predominance}d) has a chemical potential distance of 0.033~eV/atom, which is only non-zero because of how metastable phases like \ce{Mg2NCl} are calculated. Further details are in the Supporting Information (Tables S5-S13).

The chemical potential diagram boundaries are consistent with the \textit{in situ} SXRD measurements. For the Ca-system, the annotated reaction arrows in Figure \ref{fig:predominance}a correspond to the intermediates observed to form by \textit{in situ} SXRD (Figure \ref{fig:insituWSF}), and the progression from ZrN to \ce{CaZrN2} via a \ce{Ca_$x$Zr_{2-$x$}N2} solid solution (Figure \ref{fig:wsf_v_lattice}): arrow (i) represents Equation \ref{eq:ZrCl4_ZrCl3}, arrow (ii) Equation \ref{eq:ZrCl3_Zr6NCl15}, arrow (iii) Equation \ref{eq:Zr6NCl15_ZrN}, and arrow (iv) Equation \ref{eq:ZrN_CaZrN2} (or an alternative hypothesis, Equations S1, S2). 
Similarly, the chemical potential diagram for the Mg-system also matches the observed behavior of the \ce{Mg2NCl + ZrCl4} reaction: arrow (v) represents Equation \ref{eq:Mg2NCl_ZrNCl} and arrow (vi) Equation \ref{eq:ZrNCl_MgZrN2}.

This mechanistic guidance provided the key insight that ultimately led to our successful syntheses. As the Ca-Zr-N-Cl chemical potential diagram (Figure \ref{fig:predominance}a, c) shows that \ce{ZrN} formation is unavoidable along the path to \ce{CaZrN2}, we changed our synthesis conditions (relative to the two-step synthesis of \ce{MgZrN2})\cite{rom2021bulk} to deal with the refractory ZrN intermediate. Specifically, we used \ce{Ca3N2} in slight excess to provide an oxidant, and we increased our reaction temperature to increase diffusion. 
We also used only one heating step, as a low temperature step would not avoid \ce{Zr^{4+}} reduction (whereas in the case of \ce{MgZrN2}, the two-step process maintained the oxidation state of \ce{Zr^{4+}})\cite{rom2021bulk}.
These changes led to the successful syntheses shown in Figure \ref{fig:main_CaZrN2_XRD}. 

Importantly, these chemical potential diagrams can also be used to identify thermodynamically disfavored reactions, allowing researchers to quickly move beyond systems that are unlikely to work. To demonstrate this point, we also conducted metathesis syntheses and thermodynamic calculations targeting \ce{ZnZrN2} from \ce{1.2\ Zn3N2 + ZrCl4} and \ce{2.4\ Zn2NCl + ZrCl4} (Figures S19 and S20). 
\ce{ZnZrN2} has been synthesized via combinatorial sputtering of thin films\cite{woods2022roleZnZrN2} but has not yet been synthesized in bulk. 
Our \textit{in situ} SXRD experiments and thermodynamic analysis reveal that the phase is unlikely to be synthesized via metathesis without significantly increasing $\mu_\mathrm{N}$. 
The chemical potential diagrams calculated as a function of temperature show that while \ce{ZnZrN2} is predicted to be stable at room temperature, it becomes destabilized at the more synthetically relevant temperature of 300~\textdegree{}C (Figure S19). 
Instead, ZrNCl and ZrN are calculated to be the main thermodynamic products. 
\textit{In situ} SXRD measurements confirm these predictions (Figure S20). 
Although these calculations do not rule-out the possibility of \ce{ZnZrN2} forming as a kinetic product, they do show that reactions proceeding through local thermodynamic equilibrium at elevated temperature and ambient \ce{N2} pressure will not yield the targeted phase. These chemical potential diagrams therefore serve as powerful tools to accelerate materials discovery via mechanistically guided synthesis.

Several shortcomings limit the predictive ability of chemical potential diagrams. 
First, chemical potential diagrams with $>3$ dimensions may obscure certain phases: the 4-dimensional chemical potential diagrams shown here are only visualized in 3-dimensions of chemical potential ($\mu_A$, $\mu_\mathrm{Zr}$, and $\mu_\mathrm{Cl}$) with \textit{select} N-containing phases shown as intersections with the 3-dimensional space. 
Fortunately, quaternary compositional phase diagrams provide a complementary and complete, but complicated, visualization (Figure S22). 
The geometric distances for $n$-dimensional chemical spaces can also be calculated numerically using reaction network analysis\cite{mcdermott2021graph} as further discussed in the Supporting Information (Tables S5-S14).
Chemical potential diagrams also lack true kinetic information; they can only illustrate effective kinetic barriers provided by the need of a reaction to form an additional intermediate phase to establish local equilibrium\cite{neilson2023modernist}.
Lastly, they are limited by the phases in the computational databases (e.g., the Materials Project). 
These databases, while large and growing larger, are incomplete\cite{sun2019map}.
They also only contain ordered phases (not the disordered structures we observe).

\subsection{Metastability of cation disorder in \ce{CaZrN2}}

\begin{figure}[ht!]
    \centering
    \includegraphics[width =3.2 in]{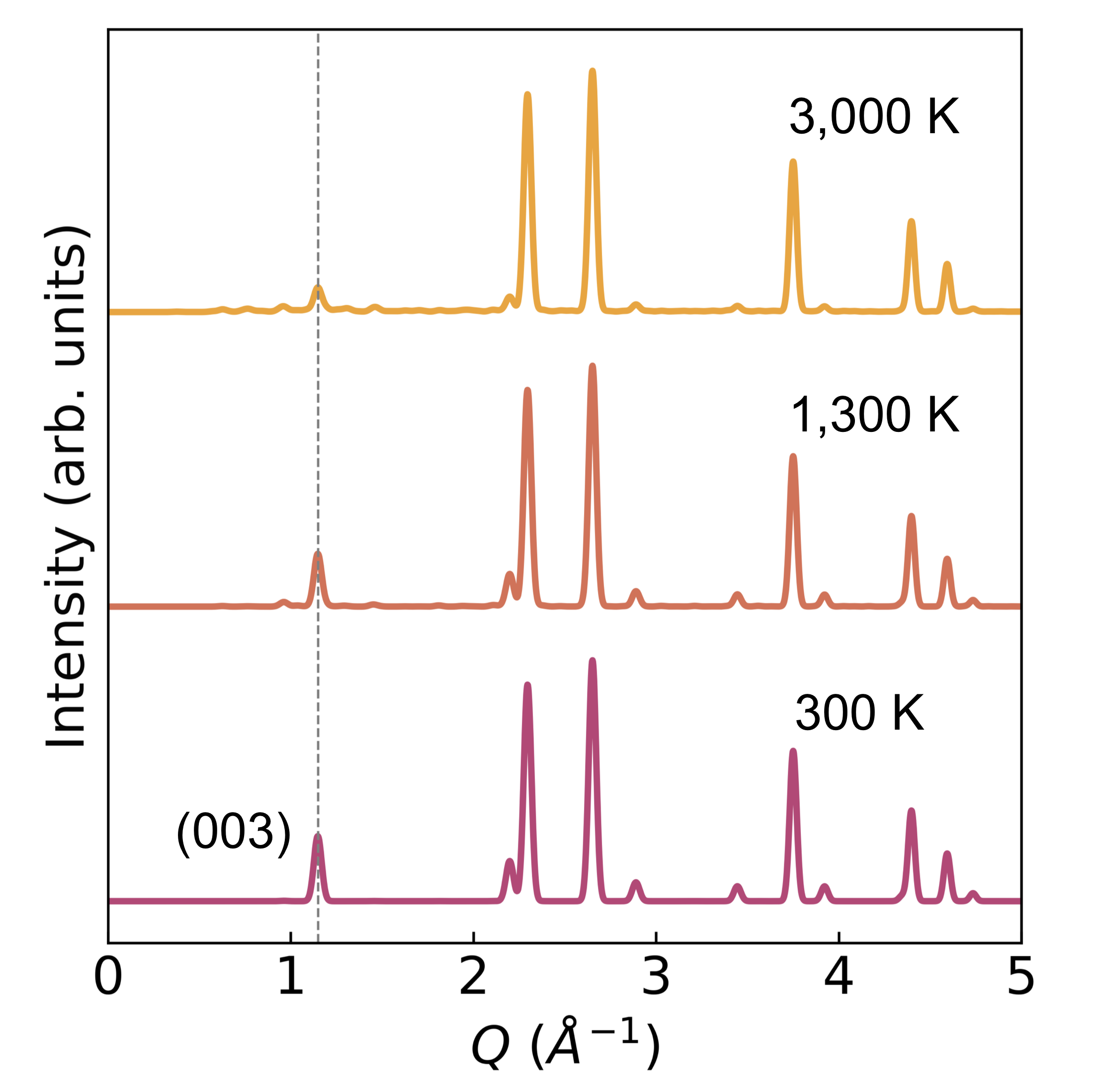}
    \caption{Simulated diffraction patterns for the thermodynamic ground state of \ce{CaZrN2}, calculated as a function of temperature from the ensemble average. The supercell reflection at $Q=1.1$~\AA{} corresponding to the (003) plane of the fully ordered ($R\bar{3}m$) structure persists even up to 3,000~K.}
    \label{fig:sim_XRD}
\end{figure}

The observed disorder in \ce{CaZrN2} is counter-intuitive.
The large contrast in atomic radii between \ce{Ca^{2+}} and \ce{Zr^{4+}} (1.00~\AA{} and 0.72~\AA{}, respectively)\cite{shannon1969effective} should favor cation ordering as observed in \ce{CaTiN2}, \ce{SrZrN2}, and \ce{BaZrN2}.\cite{li2017highCaTiN2, gregory1996synthesisSrZrN2_SrHfN2, gregory1998synthesis_AMN2_incl_BaZrN2} Indeed, through the use of DFT and cluster expansion, we found that the ground state configuration is strongly favored at 1,300~K (Figure S21). If \ce{CaZrN2} were in thermodynamic equilibrium, there would be an insignificant degree of cation disorder. 

To relate these predictions to experimental observations, we simulated PXRD patterns across all 24-atom configurations and ensemble averaged them for the temperatures of interest (Figure \ref{fig:sim_XRD}).
As expected for increasing disorder, the reflection from the (003) plane in the ordered structure (arising from cation layering) decays in intensity with increasing temperature.
Temperatures generated from DFT can be offset a couple hundred degrees from experiment\cite{novick2022mixing}, but the range from 300 to 3,000~K safely encompasses the synthesis temperature of 1,300K. 
The (003) peak in the simulated patterns is significant even at 3,000~K, suggesting that the ordering in \ce{CaZrN2} would be observed in the XRD pattern if the system were fully in thermodynamic equilibrium. 

The SXRD pattern of the experimentally synthesized \ce{CaZrN2} does not show this low-angle peak (Figure \ref{fig:main_CaZrN2_XRD}), indicating that this cation-disordered \ce{CaZrN2} is far from thermodynamic equilibrium. This hypothesis can be supported due to the short synthesis times relative to the slow rate of diffusion in nitrides.\cite{schnepf2020utilizing} Here we demonstrate an instance of a short metathesis reaction kinetically trapping disorder in a ternary nitride. Further annealing or higher synthesis temperatures could produce cation ordering in \ce{CaZrN2}, as seen in \ce{ZnGeN2} and \ce{ZnGeP2}.\cite{blanton2017characterizationOrderingZnGeN2, schnepf2019disorderZnGeP2}

This finding highlights that accounting for metastability is an area of further growth for predictive synthesis. 
As with thin film sputtering,\cite{woods2022roleZnZrN2} bulk syntheses involve myriad thermodynamic and kinetic factors that influence final product formation. 
While computational techniques to account for temperature effects on enthalpy have been developed,\cite{bartel_physical_2018} entropic contributions (e.g., compositional and configurational disorder) are not yet included in these chemical potential diagrams as they are computationally expensive to calculate.
This gap motivates further work aligning computational and synthetic techniques to bring about a truly predictive synthesis paradigm. 

\section{Conclusion}
We report the discovery of two new ternary nitrides, \ce{CaZrN2} and \ce{CaHfN2}, via mechanistically-guided metathesis synthesis of bulk powders. 
These phases crystallize in the cation-disordered RS structure ($Fm\overline{3}m$), as opposed to the computationally-predicted \ce{$\alpha$-NaFeO2} structure type ($R\bar{3}m$). 
\textit{In situ} synchrotron powder X-ray diffraction analysis shows how \ce{Zr^{3+}} intermediates form early on in the reaction process, which rationalizes why excess of the \ce{Ca3N2} reagent is synthetically necessary to generate a higher chemical potential of nitrogen. 
These findings stand in contrast to our prior synthesis of \ce{MgZrN2}, which proceeds stoichiometrically from \ce{2\ Mg2NCl + ZrCl4} to \ce{MgZrN2 + 2\ MgCl2} via a ZrNCl intermediate, as to avoid any \ce{N2}(g) linked redox chemistry. 
The observed synthetic pathways match well with the predictions generated from chemical potential diagrams. 
Additional thermodynamic calculations show that the observed $Fm\bar{3}m$ structure is metastable with respect to the ordered $R\bar{3}m$ structure, suggesting that the $R\bar{3}m$ structure may be achievable under different synthetic conditions. 
In sum, these findings demonstrate a generalizable strategy for predicting and conducting metathesis reactions with the goal of synthesizing new ternary nitrides: combine \textit{in situ} SXRD studies with thermodynamic calculations for mechanistically-guided materials chemistry.

\section{Methods}
\subsection{Synthesis}
\textit{Caution: \ce{N2} gas formation within sealed ampules can cause explosions at high temperatures, and appropriate precautions should be taken to prevent equipment damage and unsafe conditions.
}
All reagents are moisture-sensitive, and were therefore handled in an argon-filled glovebox unless otherwise noted (\ce{O2} $<$ 0.1 ppm, \ce{H2O} $<$ 0.5 ppm).
\ce{Ca3N2} (Chem Cruz or Alfa Aesar, $>$98~\% metals basis), \ce{Mg3N2} (Alfa Aesar, $>$98\% metals basis), and \ce{MgCl2} (Sigma Aldrich, anhydrous, $>$98\% metals basis) were used as received. \ce{ZrCl4} (Acros, 98\%) and \ce{HfCl4}  (Sigma Aldrich, 99\% metals basis, except for Zr, with $<2.7$\%) were purified by heating approximately 4~g in a sealed quartz ampule (10~mm inner diameter, 12~mm outer diameter, ca. 30~cm long) in a 3-zone horizontal tube furnace to transport \ce{ZrCl4} from the hot zone (400~\textdegree{}C) to the colder zone (300~\textdegree{}C), leaving behind less volatile, oxide-based impurities (e.g., \ce{ZrO2}).
\ce{CaCl2} (anhydrous, Alfa Aesar, 98\%) was dried under flowing Ar for 24~h at 300~\textdegree{}C. 

\ce{Ca2NCl} was synthesized following a method adapted from literature\cite{hadenfeldt_darstellung_1987}.
Stoichiometric amounts of \ce{Ca3N2} and \ce{CaCl2} were combined in an agate mortar and pestle and ground into a homogeneous pink powder (ca. 2~g). 
The powder was then cold pressed (P $\approx300$ MPa) into a dense pellet (diameter = 0.25~in) and placed in a niobium tube (10~cm long, 0.375~inch outer diameter, 0.015~inch wall thickness). Prior to use, the surface oxide of the niobium was removed by scrubbing with an abrasive scour pad in the glovebox. The ends of the tube were crimped down using the arbor press. This ampule was then brought out of the glovebox and quickly sealed under an argon atmosphere via arc-melting of the crimped ends of the niobium. The sealed metal ampule was then sealed in a quartz ampule under vacuum and heated at 5~\textdegree{}C/min in a muffle furnace to a set point of 740~\textdegree{}C. The reaction was held at temperature for 50~h, and then allowed to cool to room temperature before opening the ampule in the glovebox. 

\ce{Mg2NCl} was synthesized following a method adapted from literature\cite{li2015synthesis}, as we described previously\cite{rom2021bulk}. 
Stoichiometric amounts of \ce{Mg3N2} (ca. 3~g) and \ce{MgCl2} (ca. 3~g) were combined in an agate mortar and pestle and ground into a homogeneous tan powder. 
The powder was then cold pressed ($P\approx80$~MPa) into a dense pellet (diameter = 0.5~in) and placed in a quartz ampule (14 mm i.d., 16 mm o.d.). 
This ampule was then brought out of the glovebox and quickly sealed under vacuum ($\leq20$~mTorr, as determined by a Pirani gauge) by using an oxygen/methane torch. The sealed ampule (ca. 15 cm$^3$ internal volume) was then heated at 10~\textdegree{}C/min in a muffle furnace to a set point of 550~\textdegree{}C. The reaction was held at temperature for 5 days and then allowed to cool to room temperature before opening the ampule in the glovebox.

Reaction mixtures targeting \ce{CaZrN2} and \ce{CaHfN2} were prepared by combining the desired reagents in specific mole ratios (e.g., \ce{1.17\ Ca3N2 + ZrCl4}, as specified in the text) and homogenizing with an agate mortar and pestle. 
Reaction scales ranged from ca. 50~mg to ca. 500~mg total charge of reactants.
Reaction mixtures were loaded into crucibles as loose powders or as 0.25~inch diameter pellets (cold pressed at 300~MPa) and sealed under vacuum in quartz ampules ($<30$~mTorr). 
Many syntheses were conducted using alumina crucibles, but optimized syntheses used homemade stainless steel crucibles to minimize oxygen contamination.
These steel crucibles were made from stainless steel tubes (3/8~in outer diameter, 0.020~in wall thickness, TP-304L), welded closed on one end and left open on the other. 
The optimized synthesis involved pelletized reaction mixtures (ca. 500~mg) held in the steel crucibles nested inside quartz ampules along with a 1~g graphite rod as an oxygen getter, physically separated from the pellet. 
These ampules were transferred from the glovebox to a vacuum manifold using a custom air-free transfer valve, and sealed flame-sealed under vacuum ($<30$~mTorr) using an oxygen/methane torch. 
Optimized reactions were heated in a muffle furnace at $+5$~\textdegree{}C/min to the specified dwell temperature, allowed to thermally equilibrate for 10~min, and then air-quenched by removing the ampule from the furnace and placing it in an insulating brick holder on the benchtop. 
Other synthesis conditions are specified in the text where relevant. 
The products are moisture sensitive. 
To remove byproduct \ce{CaCl2} from the target compounds, samples were washed with anhydrous methanol (dried over molecular sieves for at least five days)\cite{williams2010dryingSolvents} in an argon glovebox.

\subsection{X-ray diffraction experiments}
The products of all reactions were characterized by powder X-ray diffraction (PXRD). PXRD measurements were performed by using a Bruker DaVinci diffractometer with Cu K$\alpha$ X-ray radiation. Prior to collecting PXRD, silicon powder was ground in with the reaction products as an internal standard for lattice parameters. All samples were prepared for PXRD from within the glovebox by placing powder on off-axis cut silicon single crystal wafers to reduce the background and then covered with polyimide tape to slow exposure to the atmosphere.

High-resolution synchrotron powder X-ray diffraction (SXRD) measurements on select samples were conducted via the mail-in program at the 11-BM-B end station of the Advanced Photon Source at Argonne National Laboratory\cite{wang_dedicated_2008}. Samples were prepared loading reaction products (from optimized synthetic conditions) into extruded quartz capillaries (0.69 mm inner diameter, 0.7 mm outer diameter), which were subsequently nested inside kapton capillaries for measurement. Data were collected at $\lambda \approx 0.459$\AA{}. SXRD patterns are shown in $Q$-space to account for small difference in wavelength between measurements ($Q  = 4\pi\sin(\theta)/\lambda$).  

\textit{In situ} SXRD were conducted at the 17-BM-B end station of the Advanced Photon Source at Argonne National Laboratory ($\lambda =  0.24101$~\AA). 
Samples were prepared by loading reaction mixtures into extruded quartz capillaries (0.9~mm inner diameter, 1.1~mm outer diameter) and flame-sealing under vacuum ($<30$~mTorr). 
Capillaries were loaded into a flow-cell apparatus\cite{chupas2008versatile} and heated at 5~\textdegree{}C/min to the specified temperature. \textit{Caution: \ce{N2} gas formation within the capillary can cause it to break at high temperatures.}  
Diffraction pattern images were collected using a PerkinElmer plate detector positioned 700~mm away from the sample. Images were collected every 30~s by summing 20 exposures of 0.5~s each, followed by 20~s of deadtime. 
Images collected from the plate detector were radially integrated using GSAS-II and calibrated using a silicon standard.

\subsection{X-ray diffraction analysis}
Quantitative phase analysis of PXRD and \textit{in situ} SXRD data was conducted using the Rietveld method as implemented in TOPAS v6\cite{coelho2018topas}.
For laboratory-diffraction PXRD experiments (Bruker system, Cu K$\alpha$ radiation), the sample displacement was first refined against the silicon standard.
The sample displacement was then fixed, and the relevant phases were then refined.
For each phase, lattice parameters, size broadening, and atomic thermal parameters were refined. 
A 10-term polynomial was used to fit the background. 

High-resolution SXRD datasets were analyzed in a similar way to identify the cation occupancy for the \ce{Ca_$x$ $M$_{$2-x$}N2} phase in each sample ($M$ = Zr, Hf).
Cubic ZrN ($Fm\bar{3}m$) was used as a starting model, with Ca and Zr or Hf on the cation site with 50\% occupancy each.
Cation occupancy ($x$) was constrained to equal anion occupancy (i.e., Ca + $M$ = N) during refinements. 
The anion site was fixed at full occupancy. 
The atomic displacement parameters were refined isotropically.
Other relevant phases were also added to the model, and their lattice parameters, size broadening (Lorentzian), and atomic thermal parameters were refined. 
A broad background peak from the quartz capillary was modeled with a Lorentzian function, and the remaining background was modeled with a 10-term polynomial. 
The datasets were collected without an internal standard, and sample displacement was not refined during analysis. 

Due to the number and positional overlap of intermediates during the sequential refinements, several variables were fixed to better compare phase fractions. 
Thermal displacement parameters were fixed at 1~\AA{}$^2$ for each phase to better account for changes in peak intensity during the reaction. 
Atoms were fixed at full occupancy for each site.
RS phases were modeled as ZrN (i.e., metal site mixing was not refined). 
Crystalline size domain was fixed at 200~nm (as modeled using a Lorentzian polynomial) for most phases, except the \ce{ZrNCl} and RS phases, which exhibited noticeable peak broadening. 
\ce{ZrNCl} was refined using thermal displacement parameters were fixed at 20~\AA{}$^2$ and anisotropic size broadening.
Size broadening for RS phases was manually refined from the final scan in the sequence and fixed for all other scans (crystallite size ca.~7~nm).
In order to compare the relative fractions of phases determined from Rietveld calculations, a weighted scale factor (W.S.F.) is defined as $Q_p = S_p \cdot V_p \cdot M_p$ where $Q_p$ = weighted scale factor of phase $p$, $S_p$ is scale factor calculated from Rietveld, $V_p$ is the volume of the unit cell, and $M_p$ is the atomic mass of the unit cell. 
It should be noted that we omit the Brindley coefficient for microabsorption correction in our calculation of W.S.F. due to the unreliable refinement of particle sizes for individual phases. 
Amorphous material and product lost as vapor are not accounted for in the sequential refinement, hence we use the W.S.F. instead of relative wt\% or mol\%. 
We reference all phases by their nominal stoichiometric formula; however, the actual chemical formula may be distinct from the written formula as XRD data alone cannot typically resolve nonstoichiometric compounds.

\subsection{Thermodynamic analysis}
Thermodynamic calculations were performed using DFT-computed values and tools from the Materials Project (v2022.10.28) and pymatgen (v2023.9.10)\cite{jain2013commentaryMaterialsProject, ong2013pymatgen}.
Reaction energies were calculated from the Gibbs formation energies as a function of temperature, using the method of Bartel, et al.\cite{bartel_physical_2018} Chemical potential diagrams were generated in the method of Yokokawa\cite{yokokawa1999generalizedChemicalPotentialDiagrams} as detailed by Todd, et al.\cite{todd2021selectivityPredominanceDiagrams} and implemented in pymatgen. \ce{Mg2NCl} is calculated to be +0.02~eV/atom metastable relative to the binaries, which is within DFT error of the convex hull. We therefore applied a small correction to bring the phase to the hull, allowing for visualization in the chemical potential diagrams. All other phases shown in Figures \ref{fig:predominance} and S21-S23 are calculated to be stable.

\subsection{Thermodynamic analysis of disorder in \ce{CaZrN2}}
To produce a training set for the cluster expansion, a number of DFT calculations were conducted using VASP \cite{kresse_CMS:1996}. The PBE functional was used within the projector-augmented wave method \cite{bloechl_PRB:1994}. A planewave cutoff of 400 eV was used. K-point densities were used such that the total energies were converged within 1 meV/atom. All structural degrees of freedom were allowed to be optimized within the structural relaxation (ie. volume, cell shape, atom positions). Structures were generated using the pylada software \cite{d2010pylada}. To produce symmetry-inequivalent supercells, Hermite Normal Form transformation matrices were constructed using an algorithm developed by Hart and Forcade \cite{hart2008algorithm}. The transformation matrices were applied to the RS primitive cell to produce supercells with varying dimensions. The supercells were then decorated with different atomic orderings to make varying \ce{CaZrN2} configurations.

The Alloy Theoretic Automated Toolkit (ATAT) was employed to fit a cluster expansion \cite{van2002alloy,van2002automating}. To ensure that the cluster expansion was properly representing the highly ordered configurations, we exhaustively enumerated all 8-atom and 12-atom \ce{CaZrN2} configurations and included them in the training set. In total, the training set was made up of 70 symmetry inequivalent \ce{CaZrN2} configurations, ranging in size from 8 to 24-atoms. A cross validation score of 10.4 meV/atom was obtained. Considering that the training set had an energy range of 324 meV/atom, we deemed a 10.4 meV/atom cross validation score to be sufficiently small. The energies of all 37,883 24-atom \ce{CaZrN2} configurations were predicted using the fitted cluster expansion. 

To ensemble average the XRD patterns, first the partition function, $Z$, was calculated using all 37,883 24-atom configurations \begin{equation}\label{parition function}
    Z(T)=\sum_i^{37,883} e^{-E_i/k_{B}T}.
\end{equation}  
Here, $i$ is the index of the configuration, $E_i$ is the total energy of the $i$th configuration, $k_B$ is Boltzmann's constant, and T is the absolute temperature. The temperature-dependent probability of each configuration, $P_i(T)$, was then determined using the following equation 
\begin{equation}
    P_i(T) = \frac{e^{-E_i/k_{B}T}}{Z(T)}.
\end{equation} 
Finally, the intensity of the ensemble-averaged XRD, $I$, was calculated as a function of temperature and wavenumber, $Q$: 
\begin{equation}
    I(Q,T)=\sum_i^n (I_i(Q) \cdot P_i(T)).
\end{equation}
For a justification of ensemble averaging XRD patterns in the above way, see Jones et.\,al \cite{jones2020glassy}. The powder XRD pattern for each structure was generated with a Cu K$_{\alpha}$ wavelength using pymatgen\cite{ong2013pymatgen}.

\section{Acknowledgements}
The authors thank Paul Todd and Akira Miura for helpful discussions.
This work was supported primarily by the National Science Foundation (DMR-2210780). 
The authors thank the Analytical Resources Core at Colorado State University for instrument access and training (RRID: SCR\_021758). 
Use of the Advanced Photon Source at Argonne National Laboratory was supported by the U. S.\ Department of Energy, Office of Science, Office of Basic Energy Sciences, under Contract No.\ DE-AC02-06CH11357. 
Work conducted at Lawrence Berkeley National Laboratory was supported as part of GENESIS: A Next Generation Synthesis Center, an Energy Frontier Research Center funded by the U.S. Department of Energy, Office of Science, Basic Energy Sciences under Award Number DE-SC0019212. 
This research used resources of the National Energy Research Scientific Computing Center (NERSC), a U.S. Department of Energy Office of Science User Facility operated under Contract No. DE-AC02-05CH11231.
This work was authored in part at the National Renewable Energy Laboratory (NREL), operated by Alliance for Sustainable Energy, LLC, for the U.S. Department of Energy (DOE) under Contract No.\ DE-AC36-08GO28308. Funding for work conducted at NREL was provided by DOE Basic Energy Sciences Early Career Award “Kinetic Synthesis of Metastable Nitrides”. The views expressed in the article do not necessarily represent the views of the DOE or the U.S. Government. The U.S. Government retains and the publisher, by accepting the article for publication, acknowledges that the U.S. Government retains a nonexclusive, paid-up, irrevocable, worldwide license to publish or reproduce the published form of this work, or allow others to do so, for U.S.\ Government purposes.

\section{Author Contributions}
C.L.R. and J.R.N. conceptualized the project. 
C.L.R., E.N.S., B.C.M., and G.T.T. conducted syntheses and measured PXRD. 
D.C.A. conducted optical property measurements. 
J.R.G. conducted X-ray photoelectron spectroscopy measurements with support from A.L.P.
The remote experiments were set up by beamline scientist A.A.Y., enabling C.L.R. to remotely control the data collection and conduct analysis.
R.C.M. and C.L.R. conducted scanning electron microscopy and energy dispersive X-ray spectroscopy measurements.
C.L.R. and M.J.M. conducted thermodynamic reaction calculations with support from K.A.P. and J.R.N.
A.N. conducted thermodynamic modeling of cation disorder with support from E.T. and V.S., 
A.L.P., A.Z., and J.R.N. supervised and supported the project. 

\begin{tocentry}
\begin{center}
\includegraphics[width = 3.25 in]{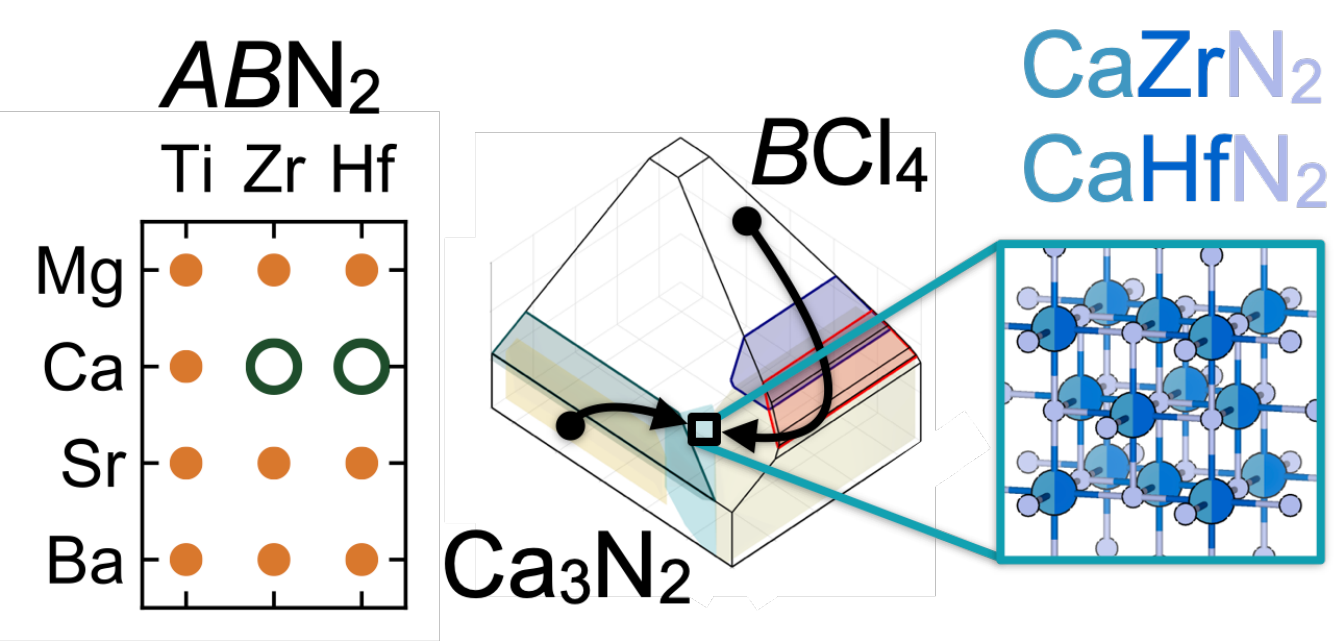}
\end{center}
\end{tocentry}

\bibliography{main} 

\end{document}